\documentstyle{article}
\begin{document}
\title{ Spatial curvature effects on molecular transport by diffusion }
\author{J.Balakrishnan \thanks{E-mail : janaki@serc.iisc.ernet.in, 
~janaki@hve.iisc.ernet.in } \\
Department of High Voltage Engineering, Indian Institute of Science, \\ 
Bangalore -- 560 012, India.  }
\date{\mbox{\ }}
\maketitle
\vspace{1.5cm}
\begin{flushright}
P.A.C.S. numbers ~~: ~~87.10.+e~,~02.90.+p ~~~~~
\end{flushright}
\vspace{1.5cm} 
\begin{abstract} 

For a substance diffusing on a curved surface, we obtain an explicit 
relation valid for very small values of the time, between the local 
concentration, the diffusion 
coefficient, the intrinsic spatial curvature and the time. We recover 
the 
known solution of Fick's law of diffusion in the flat space limit. 
In the biological context, this result would be useful in understanding 
the variations in the diffusion rates of integral proteins and other 
molecules on membranes.
\end{abstract}
\newpage

\subsection*{1. ~Introduction}

Transport of enzymes, charged ions and metabolic substances within 
biological cells and tissues and across cell membranes is one of the 
major processes which sustains and guides life. Indeed, extracellular 
and intracellular transport of substances can well be considered to be 
the most important and pervasive among all the life-supporting 
biological 
activities. \\ 

Molecular transport across cell membranes by passive diffusion or in 
accordance with Fick's law is a well-studied area. 
However in the available literature on the subject, no mention has been 
made of how the local curvature of the cell plays a part, if at all, in 
this process. Molecules released at a specific location on the cell 
surface 
or on the nuclear membrane diffuse along the curved membrane surface to 
another location. \\ 
It is known that thermal agitation permits lateral diffusion of 
phospholipid 
and glycolipid molecules within a leaflet of planar phospholipid 
bilayers of 
biological membranes. A lipid molecule can diffuse 
several micrometres per second at a temperature of 37 degrees C. It has 
also 
been established experimentally that many important proteins freely 
float 
within the plane of the membrane.\\  
Measurements have shown that the rates of diffusion of proteins in 
biomembranes are considerably lower than those seen in artificial 
membranes [1,2]. The physical structure and the dynamical changes 
occurring on a membrane surface would well be expected to play an 
important role in determining the lateral mobility of molecules on its 
surface. 
  
The metabolism and synthesis of fatty acids and phospholipids occur in 
the 
smooth endoplasmic reticulum, and the rough endoplasmic reticulum is a 
site 
of protein synthesis. It is well known that in many cells these 
extensively 
curved and folded membrane vesicles are continuous with the nuclear and 
cell membranes.  In the cytosol also, these folds distort the 
homogeneity 
in the spatial distribution of the cytosolic fluid. \\
   
Transport of a substance by diffusion should therefore be described by a 
corrected form of Fick's law, modified to take into account the local 
curvature of the surface through which it moves. \\  

In this paper we discuss how to take care of curvature effects and also 
give 
for transient phenomena, an explicit expression relating the 
concentration 
of the diffusing substance, the intrinsic spatial curvature experienced 
by 
it, the diffusion coefficient and the time.\\ 

\subsection*{2. ~Diffusion on curved surfaces}

Consider diffusion of a substance described by its concentration 
$C(x,t)$ 
from a spatial point $x$  where it has been released on the cell, to 
another point $x'$. For a particular time slice, the line element $ds$ 
between each pair of neighbouring points on the spatial surface is given 
by:\\ 

\begin{equation} 
ds^2 = \sum_{i,j=1}^n g_{ij}(x) dx^i dx^j 
\end{equation} 

where $dx^k$ denote the coordinate differences between neighbouring 
points, 
$n$ is the spatial dimension and $g_{ij}$ denotes the metric.  We choose 
to 
work with a Riemannian signature for the metric. \\ 
The usual form of Fick's law relates the current density or the flux of 
material per unit area, $j(x,t)$ to its concentration gradient in flat 
space :\\ 
\begin{equation} 
j_i(x,t) = -D\partial_i C(x,t) 
\end{equation} 
where $D$ denotes the diffusion coefficient, $\partial_i$ denotes the 
gradient 
operator, and $C(x,t)$ is the field variable denoting the concentration. 
    It is assumed here that the diffusion coefficient is independent of 
the  concentration of the diffusing substance.\\ 
In curved space, while formulating the problem, one must incorporate the 
effects of the intrinsic spatial curvature of the surface on which the 
substance is diffusing.\\
We make the simplifying assumption that in the infinitesimal 
neighbourhood of any point, the diffusion properties are the same in all 
directions and that $D$ does not depend upon the position and the 
concentration of the diffusing material.  

Transport of the substance by diffusion into and out of the invariant 
volume 
element ${\sqrt g} d^nx$ surrounding the point $x$ is given by the 
conservation equation :\\ 
\begin{equation} 
\frac{\partial C(x,t)}{\partial t} = - \nabla_i j^i(x,t) 
\end{equation} 
where $\nabla_i$  denotes the covariant derivative and includes the  
Christoffel connection ${\Gamma^k}_{il}$ ~:
$$\nabla_i j^k = \partial_ij^k + {\Gamma^k}_{il}j^l ~~~,$$ 
and we have considered a {\em parametric} dependence of $C$ on the 
time $t$. \\
Performing a covariant differentiation of (2) with respect to $x$ and 
substituting (3) in it we then get the correct form of Fick's second law 
of diffusion : \\ 
\begin{equation} 
\frac{\partial C(x,t)}{\partial t} = - D \Box C(x,t) 
\end{equation} 
where we have used $\Box$ to denote the n-dimensional Laplace-Beltrami 
operator. For flat 3-dimensional space, $\Box$ reduces to the usual 
3-dimensional Laplacian.\\ 
It has been shown in [3] that at least in mitochondrial inner membranes, 
the diffusion coefficient $D$ of intramembrane particles shows an 
inverse correlation with their concentration, implying that the proper 
form of Fick's law reflecting the concentration dependence of $D$ should 
be studied, rather than equation (4). 
In our work however, we consider only the simplest form of Fick's law 
with a concentration-independent diffusion coefficient, in order to see 
how far just the spatial curvature effects could modify the known 
result.\\
It becomes particularly interesting to learn about the configuration of 
the 
released substance during the initial infinitesimal time intervals to 
see how the intrinsic curvature of the cell would influence diffusive 
transport on the membrane surface and hence its configuration at later 
times. \\  
In order to solve (4), we rescale the time parameter by: $t \rightarrow 
Dt$ 
 so that (4) now reads:\\
\begin{equation} 
\frac{\partial C(x,t)}{\partial t} =  \Box C(x,t) - \eta C(x,t)  
\end{equation} 
and the parameter $t$ now has the dimensions of length squared. We have 
introduced a drag term $\eta C$ with $\eta >0$  which can be thought to 
account for negative concentration changes due to possible frictional 
effects on the motion of the molecules. We have introduced it here just 
for the sake of mathematical convenience and at the end of the 
calculations 
it can be set to zero. \\ 
In the actual physical situation, of course, the drag term is very much 
present and gets contributions from the drag arising from the 
pericellular 
matrix viscosity, from steric effects, and from transient binding to 
relatively immobile structures [1]. Also in the actual situation, the 
$\eta$ term is not constant and has a coupling with the concentration 
gradient. We have however restricted ourselves to $\eta= 0$ for the sake 
of simplicity in this paper.\\ 
We assume that the molecules diffuse freely on the surface without 
interacting or binding with any other molecules.
We write equation (5) in a point-separated form as : \\ 
\begin{equation} 
\frac{\partial C(x,x',t)}{\partial t} =  ( \Box_x - \eta ) C(x,x',t)  
\end{equation} 
where the biscalar $C(x,x',t)$ is subject to the condition :\\ 
\begin{equation} 
\lim_{x' \rightarrow 0} C(x,x',t) = C(x,t) 
\end{equation} 
and to the physical boundary condition:\\ 
\begin{equation} 
\lim_{t \rightarrow 0} C(x,x',t) = \delta(x,x') 
\end{equation} 

This enables us to get a well-defined explicit solution for $C(x,t)$ 
which is valid for small values of $t$, in terms of the spatial 
curvature.
The solution to (6) is well known [4-6]:\\ 
\begin{equation}
C(x,x',t) = \frac{1}{{(4\pi t)}^{n/2}} e^{-\eta t} 
e^{-\frac{\sigma(x,x')}{2t}} 
\Delta^{1/2}(x,x') \Omega(x,x',t) 
\end{equation} 
where the biscalar $\sigma(x,x')$ equals half the square of the geodesic 
distance between $x$ and $x'$ and $\Delta(x,x')$ is the VanVleck-Morette 
determinant :\\ 
\begin{equation} 
\Delta(x,x') = - {(g(x))}^{-1/2} \det 
[-\partial_i\partial_{j'}\sigma(x,x')]
{(g(x'))}^{-1/2} 
\end{equation} 
This is a biscalar quantity which reduces to unity in flat space. \\ 
In curved space, one can expand $\Delta(x,x')$ in a series expansion in 
powers of the curvature by working in Riemann normal coordinates $y$ 
which define a locally inertial system in the neighbourhood of the point 
$x'$. In these coordinates [7], with origin at $x'$, ~~ $\Delta(x,x') = 
(g(x))^{-1/2}$ ~, so that \\
\begin{equation} 
\Delta^{1/2}(x,x') = (g(x))^{-1/4} = 1 + 
\frac{1}{12}R_{\alpha\beta}y^\alpha y^\beta + O(y^3) 
\end{equation} 
where $x$ is regarded as a function of the Riemann normal coordinates 
$y$, such that ~~ $x \rightarrow x'$ ~ as ~ $y \rightarrow 0$.~ In the 
coincidence limit, and for our purposes, it is only the first term on 
the right hand side of (11) which is relevant for the calculations. \\
The function $\Omega(x,x',t)$ has the following series expansion in the 
coincidence limit  $x' \rightarrow x$ :\\ 
\begin{equation} 
\lim_{x' \rightarrow x} \Omega(x,x',t) = \sum_{k=0}^\infty t^k E_k(x) 
\end{equation} 
valid in the limit $t \rightarrow 0$  where $E_k(x)$ are known 
coefficients 
known in the literature as Gilkey coefficients [4-7] : \\ 
\begin{eqnarray} 
E_0 & = & I \nonumber\\ 
E_1 & = & \frac{R}{6} - \eta \nonumber\\ 
E_2 & = & \frac{1}{2}{(\frac{R}{6} - \eta )}^2 - \frac{1}{180}R_{\mu\nu} 
R^{\mu\nu} + \frac{1}{180}R_{\mu\nu\rho\sigma}R^{\mu\nu\rho\sigma} + 
\frac{1}{30}\Box R - \frac{1}{6}\Box \eta \nonumber \\ 
E_3 & = & \frac{1}{7!}{\bigl [} 18\Box^2 R + 17R_{;\mu}R^{;\mu} - 
2R_{\mu\nu;\rho}
R^{\mu\nu;\rho} - 4R_{\mu\nu;\rho}R^{\mu\rho;\nu} + 
9R_{\mu\nu\rho\sigma;\tau}R^{\mu\nu\rho\sigma;\tau} + 28R\Box R 
\nonumber\\ 
&   &  - 8R_{\mu\nu}\Box R^{\mu\nu} + 24R_{\mu\nu}{R^{\mu\rho;\nu}}_\rho 
+ 12 R_{\mu\nu\rho\sigma}\Box R^{\mu\nu\rho\sigma} + \frac{35}{9}R^3 
-\frac{14}{3}RR_{\mu\nu}R^{\mu\nu} + \frac{14}{3}RR_{\mu\nu\rho\sigma}
R^{\mu\nu\rho\sigma} \nonumber\\ 
&   & -\frac{208}{9}R_{\mu\nu}{R^\mu}_\rho R^{\nu\rho} + 
\frac{64}{3}R_{\mu\nu}
R_{\rho\sigma}R^{\mu\rho\nu\sigma} -\frac{16}{3}R_{\mu\nu}
{R^\mu}_{\rho\sigma\tau}R^{\nu\rho\sigma\tau} + 
\frac{44}{9}R_{\mu\nu\rho\sigma}R^{\mu\nu\alpha\beta}
{R^{\rho\sigma}}_{\alpha\beta} \nonumber\\ 
&   & + \frac{80}{9}R_{\mu\nu\rho\sigma}
R^{\mu\alpha\rho\beta}{{{R^\nu}_\alpha}^\sigma}_\beta {\bigr ]}  
- \frac{1}{60}\Box^2\eta + \frac{1}{12}\eta \Box \eta + 
\frac{1}{12}(\Box \eta)
\eta + \frac{1}{12}\eta_{;\mu}\eta^{;\mu} - \frac{1}{6}\eta^3 
-\frac{1}{36}R\Box \eta \nonumber\\ 
&   & - \frac{1}{90}R^{\alpha\beta}\eta_{;\alpha\beta} 
-\frac{1}{30}R_{;\mu}\eta^{;\mu} + \frac{1}{12}\eta^2R -\frac{1}{30}\eta 
R^2 
+ \frac{1}{180}\eta R_{\mu\nu}R^{\mu\nu}  
- \frac{1}{180}\eta R_{\mu\nu\rho\sigma}R^{\mu\nu\rho\sigma}  
\end{eqnarray} 
Here, $I$ denotes the identity matrix, $R$ stands for the Ricci scalar 
and 
the semicolon denotes a covariant differentiation. Although the fourth 
Gilkey 
coefficient has also been calculated in the literature, we have 
displayed above 
only terms upto third order in the Riemann curvature. \\ 
Rescaling now $t$ back to $Dt$ we obtain the solution we seek for 
diffusion of 
molecules in the presence of a drag term in n spatial dimensions for 
transient 
times :\\
\begin{equation}
C(x,t) = \frac{1}{{(4\pi Dt)}^{n/2}} e^{-\eta Dt} 
e^{-\frac{\sigma(x,0)}{2Dt}}
\Delta^{1/2}(x,0) \sum_{k=0}^\infty {(Dt)}^k E_k(x)
\end{equation}

For the standard diffusion equation without the $\eta$ term, and in flat 
space, the only Gilkey coefficient which contributes is $E_0$ and in 
this case we recover the known result:\\ 
\begin{equation} 
C(x,t) = \frac{1}{{(4\pi Dt)}^{n/2}} e^{-\frac{x^2}{4Dt}} 
\end{equation} 

In a recent paper [8], Gompper and Goos have suggested that the 
diffusion of amphiphilic molecules within a monolayer at the oil-water 
interface of the microemulsion phase in an oil-water-amphiphile mixture 
can be used to measure the average Gaussian curvature of the monolayer. 
They have considered surfaces of constant curvature. The result (14) we 
have discussed here for the concentration, is valid for surfaces of 
varying curvature also. 

In fact, from (14) it is an easy matter to get a general solution to the 
diffusion equation for $n$-dimensional spaces with arbitrary constant 
curvature $K$ for which the value of the Riemann curvature depends 
neither on the coordinate $x$ nor on the planar direction at $x$.  For 
such spaces, the Riemann curvature is given in terms of their metric 
$g_{ij}$ by \\ 
\begin{equation}
R_{ijkl} = K(g_{ik}g_{jl} - g_{il}g_{jk})    ~~~~ ~~~({\rm for}~~~ n \ge 
3)~~, ~
\end{equation} 
whence the Gilkey coefficients turn out to be :\\
\begin{eqnarray}
E_0 &=& 1 \nonumber\\
E_1 &=& \frac{n(n-1)}{6}K \nonumber\\ 
E_2 &=& \frac{n(n-1)(3n+1)}{360}K^2 \nonumber\\ 
E_3 &=& \frac{n(n-1)}{9\times 7!}\Bigl \{ 7{(n-1)}^3 (5n-1) + 
61{(n-1)}^2 + 68n + 28 \Bigr \} K^3 
\end{eqnarray} 
We then find that the expression (14) for the concentration of the 
diffusing substance has the following dependance on the Gaussian 
curvature $K$ :\\
\begin{eqnarray} 
C(x,t) &=& \frac{1}{{(4\pi Dt)}^{n/2}} e^{-\frac{x^2}{4Dt}}\Bigl ( 1 + 
\frac{n(n-1)}{6}KDt + \frac{n(n-1)(3n+1)}{360}{(KDt)}^2 \nonumber\\ 
&+& \frac{n(n-1)}{9\times 7!}\bigl [ 7{(n-1)}^3(5n-1) + 61{(n-1)}^2 + 
68n + 28 \bigr ] {(KDt)}^3 + \dots \Bigr )  
\end{eqnarray}
$K >0$ corresponds to the spherical surfaces while surfaces with $K<0$ 
correspond to hyperboloid ones --- the $K=0$ are flat Euclidean 
surfaces.\\  
It is shown in [8] that the structure of a microemulsion can be 
quantified in terms of a quantity which depends upon the Euler 
characteristic $\chi_E$ of the surface within which the amphiphile 
molecules diffuse. $\chi_E$ is obtained from the Gaussian curvature 
using the Gauss-Bonnet theorem :
\begin{equation}
\int dS K = 2\pi \chi_E 
\end{equation} 
where the integral is over a closed surface S. It should be therefore 
possible in the case of surfaces of approximately constant area, to 
express the result (18) in terms of the topological invariants 
characterizing them, after appropriately scaling them. This exercise is 
however beyond the scope of this report. Because of their enormous 
complexity, biological cells and membranes do not in general have 
isotropic and homogeneous composition, and the membrane surfaces are 
more often than not, of varying curvature, and in these situations, one 
needs to use (13) and (14) rather than (18).\\  

For the specific case of diffusion in two dimensions such as on 
membranes, the coefficients in (13) simplify considerably because in 
these dimensions both the Riemann tensor $R_{ijkl}$ and the Ricci 
curvature scalar $R$ have only one component and both the Riemann tensor 
and the Ricci tensor $R_{ij}$ can be expressed in terms of the curvature 
scalar 
$R$ :\\ 
\begin{equation} 
R_{ijkl} = \frac{1}{2}R(g_{ik}g_{jl} - g_{il}g_{jk}) 
\end{equation} 
and\\
\begin{equation}
R_{ij} = \frac{1}{2}Rg_{ij}
\end{equation} 
It must be borne in mind that in our treatment, we have regarded time as 
a 
{\em parameter} and the indices $i, j, k, l, \mu, \nu, \alpha, \beta$ 
etc. label spatial dimensions only, for we are working on a particular 
time slice at each instant of time.\\
We consider the simplest example of diffusion of a substance on the 
surface of a sphere of constant radius $r$. For a 2-sphere, the Ricci 
curvature scalar is:
$$ R = \frac{2}{r^2}  .$$ In fact, in this case, the Gaussian curvature 
$K = \frac{1}{r^2}$.  
Substituting this value of $R$ in (13), (20) and (21), or by simply 
using (17), the Gilkey coefficients reduce for this example to:
\begin{eqnarray}
E_0 &=& 1 \nonumber\\
E_1 &=& \frac{1}{3r^2} \nonumber\\ 
E_2 &=& \frac{1}{15r^4} \nonumber\\ 
E_3 &=& \frac{4}{315r^6} 
\end{eqnarray} 
giving the following result for the concentration of the diffusing 
substance of an initial unit amount, at a point distant $x$ from the 
point of its release on the surface of the sphere, at a time $t$:\\
\begin{equation} 
C(x,t) = \frac{1}{(4\pi Dt)} e^{-\frac{x^2}{4Dt}}\Bigl ( 1 + 
\frac{Dt}{3r^2} + \frac{1}{15}{(\frac{Dt}{r^2})}^2 + 
\frac{4}{315}{(\frac{Dt}{r^2})}^3 + \dots \Bigr )  
\end{equation} 

The result obtained in [8] for the mean square displacement of a 
particle diffusing on a sphere is essentially equivalent to the leading 
and next to leading order terms in (23).

In the series expansion in (14), (18) and (23), valid for very small $t$ 
values, it is assumed that the curvature terms are small in comparison 
with the flat space result. Care must be taken before applying the 
actual values of $t, D$ and $r$ to these expressions to ensure that this 
assumption is satisfied. 

For a substance having a $D$ value of $10^{-6}{\rm cm}^2{\rm s}^{-1}$  
released on the surface of a spherical cell of radius $1\mu$m , 
diffusing through a distance of $0.5 \mu$m in time 1ms, one obtains a 
calculated value of $4259.4751 \times 10^4$ per ${\rm cm}^2$ for its 
concentration, using the usual expression (15) for flat space diffusion, 
while the improved solution (23) gives an additional correction of 0.034 
per ${\rm cm}^2$ to this --  a difference of  3.4\% from the flat space 
result, and a deviation of 0.33\% from the flat space results for a time 
duration of $10^{-4}$s, while for a $D$ value of $10^{-7}{\rm cm}^2{\rm 
s}^{-1}$, the deviations from the flat space result for time durations 
of 1ms and 0.1ms are 0.33\% and 0.033\% respectively. \\
The experimentally measured values of the concentrations, of course, 
correspond to the corrected values and the curvature-corrected Fick's 
law (14), since the diffusing molecules have already traversed over the 
curved surface of the cell. However, it must be borne in mind that  the 
diffusion times measured and calculated, in fact, calibrate distances 
different from the flat space distances, when one makes a  comparison 
between diffusion rates on different cells and on membranes whose 
curvatures differ from point to point and from one another. What we 
intend to point out here is that one must remember that flat space 
methods must not be applied when one is talking about biological 
membranes and surfaces which are very curved, or even for surfaces with 
varying curvature for which one must apply the coefficients in (13). We 
have not included realistic effects such as drag terms arising from the 
viscosity of the cytosolic fluid, and we have considered also only 
lateral (2-dimensional) diffusion in this work.\\  

It is seen that curvature effects modify considerably the solution of 
the 
diffusion problem. In the biological context, it is well known that 
depending 
upon the cell type, between 30-90 percent of all integral proteins in 
the 
plasma membrane are freely mobile and among these, the lateral diffusion 
rate of a protein in an intact membrane is around 10-30 times lower than 
that 
of the same protein embedded in synthetic liposomes [1,2,9]. It has been 
suggested that this could be because the mobility of the proteins might 
be hampered by 
interactions with the rigid submembrane cytoskeleton. \\ 
In arriving at those diffusion rates, these authors considered only the 
normal form of Fick's law for flat space. Use of the correct form of 
Fick's law taking into account the varying curvatures of the membranes 
on which the protein molecules diffuse must be made when seeking to 
explain through such theories. 

\subsection*{Discussion} 

We have shown how the curvature of the surface through which molecules 
diffuse 
modify the usual form of Fick's law and the relation between the 
concentration 
of the diffusing molecules, the diffusion constant and the time. \\ 
Many intra-membrane particles are electrically charged and when they are 
subjected to an external electric field, move from their original random 
distribution to a more ordered distribution. It would be interesting to 
see 
the effect of external electromagnetic fields on molecules which are 
electrically charged, diffusing on curved surfaces.

\subsection*{Acknowledgement} 

I would like to acknowledge support from the Jawaharlal Nehru Centre for 
Advanced Scientific Research, Bangalore, during the course of this 
work.\\
 
\newpage
\subsection*{References}
\begin{enumerate}
\item F.Zhang, G.M.Lee \& K.Jacobson,  Bioessays {\bf 15}, 579 (1993) ~~ 
and references therein.
\item K.Jacobson, A.Ishihara \& R.Inman,  Ann.Rev.Physiol.{\bf 49}, 163 
(1987). 
\item A.E.Sowers \& C.R.Hackenbrock,  Biochim.Biophys.Acta {\bf 821}, 85 
(1985). 
\item P.Gilkey, J.Diff.Geom.{\bf 10}, 601 (1975).
\item B.S.DeWitt, {\em Dynamical Theory of Groups and Fields},
(Gordon \& Breach, N.York, 1965). 
\item L.Parker and D.J.Toms, Phys.Rev.{\bf D31}, 953 (1985).\\
L.Parker and D.J.Toms, Phys.Rev.{\bf D31}, 3424 (1985). 
\item L.Parker in {\em Recent Developments in Gravitation --
Cargese  1978 Lectures}, eds. M.Levy and S.Deser (Plenum Press,
 New York, 1979).
\item G.Gompper \& J.Goos, in {\em Annual Reviews of Computational 
Physics II} --ed. D.Stauffer, 101- 136. (World Scientific, Singapore 
(1995)). 
\item H.Lodish, D.Baltimore, A.Berk, S.Lawrence Zipursky, P.Matsudaira 
\& J.Darnell, 
{\em Molecular Cell Biology}, (Scientific American Books, N.York, 
1995)(3rd edition). 
\end{enumerate}

\end{document}